\documentclass[usenatbib]{mnras}

\usepackage{newtxtext,newtxmath}

\usepackage[T1]{fontenc}
\usepackage{ae,aecompl}

\usepackage{graphicx}	
\usepackage{amsmath}	
\usepackage{amssymb}	

\usepackage{longtable}  
\usepackage{environ}  
\usepackage{pdflscape}  
\usepackage{threeparttable}  
\usepackage{threeparttablex}  

\title[Flows of Local Sheet Dwarfs]{Flows of Local Sheet Dwarfs in Relation to the Council of Giants}

\author[L. M. Seaton, M. L. McCall, and N. T. McCall]{
Lucas M. Seaton, Marshall L. McCall,\thanks{E-mail: mccall@yorku.ca} and Neil T. McCall
\\
Department of Physics and Astronomy, York University, Toronto, Ontario, Canada \  M3J~1P3\\
}

\date{Accepted 2024 February 21. Received 2024 February 13; in original form 2023 August 06}

\pubyear{2024}

\begin{document}
\label{firstpage}
\pagerange{\pageref{firstpage}--\pageref{lastpage}}
\maketitle

\begin{abstract} 
The kinematics of isolated dwarf galaxies in the Local Sheet have been studied to ascertain how the Council of Giants has affected flows.   Peculiar velocities parallel to the Sheet in the frame of reference of the Council ascend steeply from negative to positive values on the near side of the Council at a heliocentric radius of 
$2.4 \pm 0.2 \, \rm Mpc$.  
{
They descend to preponderantly negative values at a radius of 
$3.9^{+0.4}_{-0.5} \, \rm Mpc$},
which is near the middle of the Council realm.  
Such behaviour is evidence for a flow field set up by the combined gravitational effects of the Local Group and Council,    
{the ascending node being where their gravitational forces balance.}
Receding dwarfs on the near side of the Council are predominantly located in the direction of M94, {although this may be a manifestation of} the limitations of sampling.  
If M94  were entirely responsible for the placement of the ascending node, then the galaxy's total mass relative to the Local Group would have to be 
{
$0.8^{+0.4}_{-0.3}$, 
the same as indicated by the orbits of satellite galaxies.} 
Rather, if the placement of the ascending node were set by matter distributed evenly in azimuth at the Council's radius, then the required total mass relative to the Local Group would have to be
{
$4^{+3}_{-2}$,
which is 30\% to 40\% lower than implied by satellite motions but still consistent within errors.} 
The mere existence of the ascending node confirms that the Council of Giants limits the gravitational reach of the Local Group.
\end{abstract}
 

\section{Introduction} \label{intro}

{
The Local Group is encircled by giants in the Local Sheet ($M_\textit{Ks} \le -22.5$) concentrated around a mean radius of 
$3.75 \, \rm Mpc$ 
from a centre which is offset from the Sun by 
$0.8 \, \rm Mpc$  
(see Figure~\ref{sheet_views}).
}
The collection is dynamically distinct from the Local Supercluster because
angular momentum vectors are organized very differently from those of galaxies beyond.  The configuration has been called the ``Council of Giants'' \citep{mcc14a}.    Although the Council may have arisen by chance, it nevertheless represents a proximate concentration of bright galaxies with an unusually high density relative to the field and simulated analogues of the Local Volume \citep{neu20a}.   As such, it may have influenced the development of the Local Group, because the gravitation of the Council limits the region of space from which the Local Group is able to pull matter.

If indeed the Council of Giants has influenced local evolution, then evidence may be found today in the peculiar motions of galaxies in the Local Sheet on either side of a boundary nearer than the Council where the gravitational forces of the Local Group and the Council balance.  Galaxies on the near side of the boundary (i.e., closer to the Local Group) should be drawn towards the Local Group and consequently should be moving towards it.  
Galaxies on the far side of the boundary but still on the near side of the Council ought to be drawn towards the Council and consequently should be moving away from the Local Group.  Galaxies on the far side of the Council should be drawn towards the Council also, but should be moving towards the Local Group.  This paper focuses on the motions of isolated dwarf galaxies in the frame of reference of the Council to determine if there are any flows that fit this pattern.  Karachentsev and collaborators have conducted many studies of the motions of galaxies in groups located in the Local Sheet \citep{kar02a, kar02b, kar02c, kar03a, kar03b, kar03c, kar05a}, but as far as is known no investigation of motions more globally has been conducted in relation to the Council of Giants.  

In \S\ref{sample}, the sample of dwarfs used to probe local flows is constructed.  Positions and peculiar velocities are established in \S\ref{measurements}.  In \S\ref{analysis}, the spatial distribution of peculiar velocities is examined and evidence for organized motions is presented.  Implications are discussed in \S\ref{discussion}, and conclusions are presented in \S\ref{conclusions}.

\section{Sample of Galaxies} \label{sample}

Many dwarf galaxies are suitable to use as test particles for studying flows within constituents of the cosmic web.   Motions that deviate from the Hubble flow incorporate responses to the gravitational field of the environment, yet the dwarfs themselves have little effect on the organization of the massive bodies establishing that field.   Furthermore, they are the bodies out of which the largest galaxies have been built as a consequence of flows.  However, some lie close to a single large galaxy, in which case peculiar motions may not be reflective of the influence of matter distributed broadly.  {Nevertheless, through careful selection, dwarfs offer the means to trace flows established by collections of galaxies.  Those suitable for tracing flows in relation to the Council of Giants must have reliable distances and velocities, be well-separated from more massive neighbours, and be legitimately considered to be members of the Local Sheet.}

{
Candidates for the study were extracted on 2023 September 22 from the 
Catalog \& Atlas of the LV Galaxies\footnote{https://www.sao.ru/lv/lvgdb/} 
(LVG; see \citealt{kai19a}), 
which is the current expanded on-line version of the Updated Nearby Galaxy Catalog \citep{kar13a}.  Attention was restricted to objects for which there were data sufficient to determine their distances from the tip of the red giant branch (TRGB) and for which the heliocentric velocity was known to an accuracy of 
$20 \, \rm km \, s^{-1}$  
or better (see \S\ref{distances_and_velocities}).   Adoption of a single reliable distance indicator ensured that all of the objects could be placed on the same distance scale.  The sample was limited to galaxies lying within 
$6.25 \, \rm Mpc$  
of the centre of the Council of Giants 
(5.4 to $7.1 \, \rm Mpc$  
from the Sun), which defines the spherical volume of space within which all giants and nearly all dwarfs are confined to the Local Sheet (see \citealt{mcc14a}).  The distance criterion limited the number of candidates for study to 
104,  
of which
102  
had the required velocity accuracy.}

{
The degree of isolation of each candidate was judged from the tidal index of gravitationally influential neighbours  \citep{kar99a}.}  The tidal index $\Theta_\textit{N}$ arising from the $N$ most significant "disturbers" in the vicinity of the dwarf is defined by the mass $M_i$ and the distance $D_i$ of each disturber $i$:
\begin{equation} \label{eq_iso}
\Theta_\textit{N} = \log \left[ \, \sum_{i=1}^N \frac{M_i}{D^3_i} \, \right] + C
\end{equation}
\noindent
Here, $C$ is a constant and $N$ is the number of disturbers considered.  In the Updated Nearby Galaxies Catalog, $C$ was selected to be 
$-10.96$,  
which yielded $\Theta_1 = 0$ for a dwarf located at the radius of the  "zero velocity sphere" of the main disturber.  Galaxies with $\Theta_\textit{N} < 0$ can be regarded as being isolated because they still retain a memory of the Hubble flow.   Higher $N$ implies stricter isolation.  
{
The sample was restricted to the
94  
candidates meeting the distance and velocity criteria for which {\it both} $\Theta_1<0$ and $\Theta_5 <0$ in the LVG.}

{
To ensure that sample galaxies would be sensitive to any tug of war between the Local Group and the Council of Giants, their altitude above or below the Local Sheet was required to be no larger than half the radius of the Council, which is 
$1.9 \, \rm Mpc$.  
This corresponds to 
2.5  
times the vertical dispersion of the isolated members of the Sheet.   A total of 
84 
candidates met this constraint.  The
10 
rejected galaxies were offset from the Sheet plane by 
$2.2 \, \rm Mpc$ 
 or more, and all but two were located near the edge of the Sheet, more than 
 $5 \, \rm Mpc$  
 from the Council centre.}

{
Rotationally supported galaxies are likely to have experienced mergers and so are not ideal objects to include in the sample.  Rather, the sample was restricted to low-mass galaxies supported primarily by pressure, what \citet{ivk19a} define to be true dwarfs.  Such galaxies have a velocity dispersion of about 
$30 \, \rm km \, s^{-1}$  
or less and an absolute magnitude in $K_s$ of around 
$-18$  
or fainter.  Accounting for uncertainties and scatter, sample galaxies were required to have a velocity dispersion of at most 
$40 \, \rm km \, s^{-1}$  
 (if measured) and an absolute magnitude in $K_s$ no brighter than 
 $-19.5$ (as derived from the LVG).  
 The former translates into an HI line width of at most 
 $94 \, \rm km \, s^{-1}$ at 50\% of peak (as recorded in the LVG).  
Of course, a rotating galaxy observed from the side may well have a line width less than this, but the absolute magnitude criterion would eliminate it.  Applying the line width and absolute magnitude constraints, the list of candidates was reduced to a final sample of
68  
galaxies, of which 
59  
have late-type classifications ($T \ge 9$).
They are listed in Table~\ref{mccall_tab1}, sorted in order of heliocentric distance projected along the plane of the Local Sheet (see \S\ref{sheet_coordinates}).}

\section{Measurements} \label{measurements}
\subsection{Distances and Velocities} \label{distances_and_velocities}

{
Modern homogeneous measurements of TRGB distances are collected in the CosmicFlows-4 Database \citep{tul23a} attached to the 
Extragalactic Distance Database\footnote{http://edd.ifa.hawaii.edu/dfirst.php} (EDD),
from which distance moduli as of 2023 October 17 were extracted for all but two of the sample galaxies.  To bring them on to the same scale as that of the Council of Giants,  reference was made to the work of \citet{fin12a}, who established TRGB distances for dwarf galaxies in the Local Sheet by homogenizing evaluations of extinction, reddening, and K-corrections for red giant stars using the same techniques used for giant galaxies in the Sheet \citep{mcc14a}.
The calibration of \citet{riz07a} was employed to link the absolute magnitude of the TRGB in $I$ to its $V - I$ colour.  \citet{fin12a} anchored the zero-point of the calibration to the maser distance of 
$7.2 \pm 0.3 \, \rm Mpc$  
 determined for NGC~4258 by \citet{her99a}.  However, \citep{mcc14a} based distances for giants upon an improved maser distance of 
 $7.60 \pm 0.23 \, \rm Mpc$ \citep{hum13a}.   
Therefore, distance moduli derived by \citet{fin12a} were updated accordingly. 
Distance moduli in CosmicFlows-4 for 50 galaxies in common proved to be systematically lower, so they were increased by 
$0.052 \, \rm mag$  
to put them on the same scale as that of Council giants.
This correction happened to lead also to a match with the distance scale adopted in the LVG, from which the TRGB distances to two of the sample galaxies 
(Leo P and AGC~749241)
were taken.  The mean uncertainty of derived distance moduli for sample galaxies is 
$0.07 \, \rm mag$.}  

{
Heliocentric radial velocities were extracted from 
HyperLeda\footnote{http://leda.univ-lyon1.fr/}
\citep{mak14a}, 
which summarizes measurements to date along with their uncertainties and sources for all sample galaxies, particularly distinguishing between radio and optical values.   Error-weighted mean values as of 2023 October 28 were adopted.  For 58  
of the 
68  
sample galaxies, the uncertainty in the velocity is 
$5 \, \rm km \, s^{-1}$  
or less.  Velocities are within $10 \, \rm km \, s^{-1}$ of those recorded in the LVG for 65 of the sample galaxies.  The largest deviation is $25 \, \rm km \, s^{-1}$, but for that galaxy the adopted velocity is within  $1\, \rm km \, s^{-1}$ of that listed in the EDD.}

{
Adopted distance moduli and velocities along with their uncertainties are given in Table~\ref{mccall_tab1}.}

\subsection{Sheet Coordinates}\label{sheet_coordinates}
The Local Sheet is an extremely flattened system, so it is logical to explore flows in a frame of reference defined by it.  Such a frame, referred to here as  the ``Rotated Sheet Coordinate System'', was established by \citet{mcc14a} by fitting a plane to the giants.  
The plane is tilted by 
$8^\circ$ 
with respect to the supergalactic plane, with its 
north pole located at supergalactic coordinates 
$(L, B) = (242^{\circ}, +82^{\circ})$.  
The Sun is offset by 
$0.13 \, \rm Mpc$ north  
of the plane.  The coordinate system's origin is the perpendicular projection of the Sun onto the plane.  The positive Z-axis is perpendicular northward from the Sheet and the positive Y-axis is rotated about the Z-axis by 
$107^\circ$ clockwise  
from the direction of the supergalactic longitude of the pole.  

\begin{figure} 
\includegraphics[keepaspectratio=true,width=8.5cm]{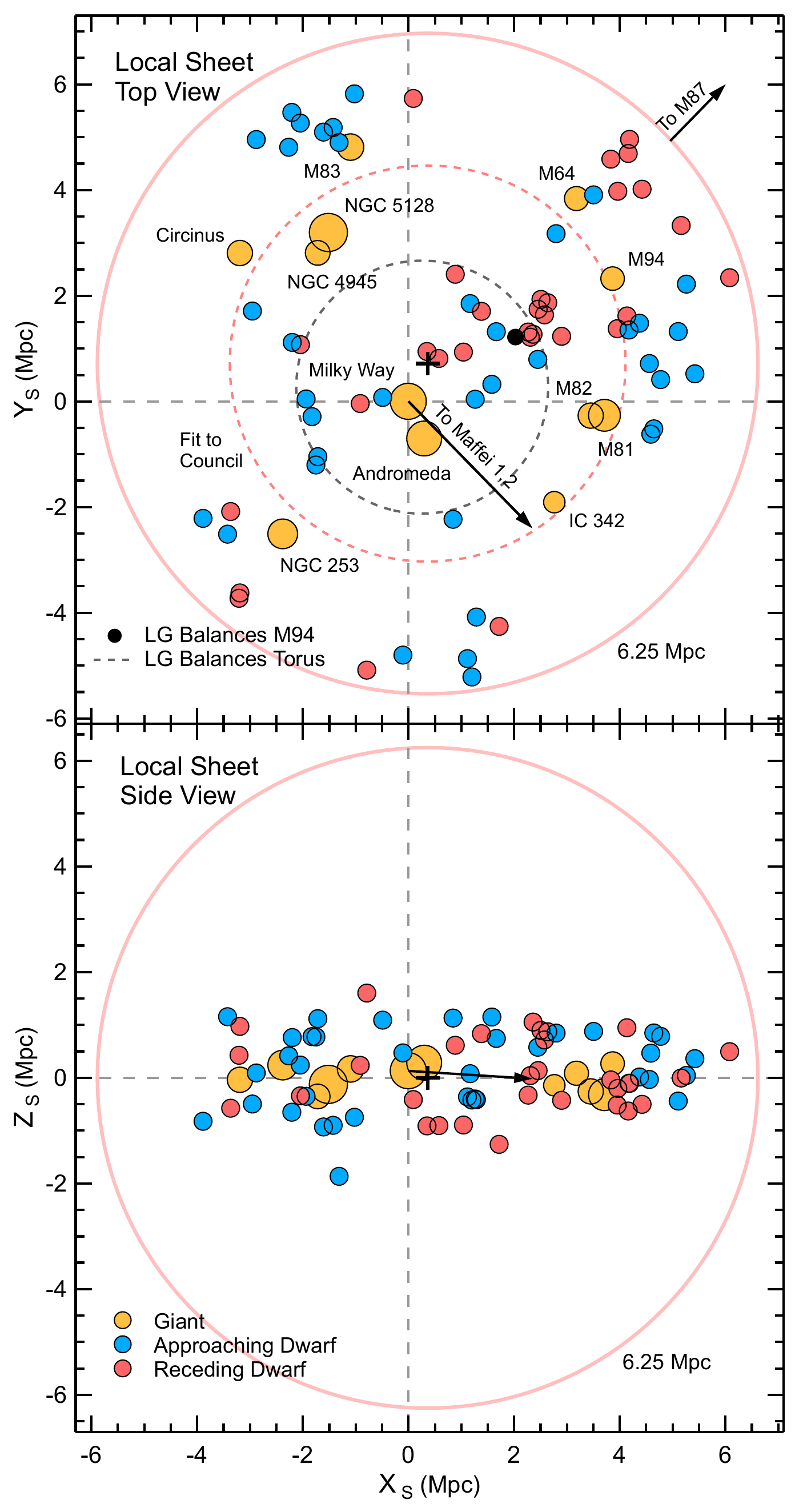}
\caption{\label{sheet_views}
Top and side views of the Local Sheet (upper and lower panel, respectively) as depicted by Rotated Sheet Coordinates.  {The dashed horizontal and vertical lines are the coordinate axes.  A solid pink circle centred on a black plus delimits the spherical volume of radius $6.25 \, \rm Mpc$ within which galaxies are sampled}.  Each giant galaxy is marked with  a filled yellow circle that is sized in proportion to the cube root of the luminosity in $K_s$.  A solid black arrow {emanating from the Milky Way} shows the direction of the centroid of Maffei 1 and 2, the distance for which is uncertain (the tip is at the location specified by \citealt{mcc14a}).   Isolated dwarf galaxies are represented by small circles of a fixed size (not to scale).  Colours are based upon the projected peculiar velocity:  blue for approaching and red for receding.   {In the top view, the dashed pink circle shows the fit to the Council of Giants, the black plus being its centre.  Also, a second black arrow labeled "To M87" shows the direction to the centre of the Virgo Cluster as seen from the Sun.  A solid black circle marks where the gravitational pull of the Local Group balances that of M94 if the mass of M94 is 0.80 times that of the Local Group.  A dashed black ellipse delineates where the pull of the Local Group balances the net attraction of a thin uniform torus with the radius of the Council and a mass 4.3 times that of the Local Group.}}
\end{figure}

Cartesian coordinates in the Rotated Sheet Coordinate System for each sample galaxy were determined from the supergalactic coordinates {(from HyperLeda)} and adopted distance.  They are provided in Table~\ref{mccall_tab1}.  Also given in the Table is the component of the heliocentric distance parallel to the Sheet, which will henceforth be designated as the {\it projected distance}.
Figure~\ref{sheet_views} illustrates how the dwarfs are distributed with respect to the giants 
of the Local Sheet as seen from above and from the side.  
With Cartesian coordinates in the Sheet  system specified by  
$(X_\textit{S}, Y_\textit{S}, Z_\textit{S})$,  
the centroid of the Local Group (defined as the  $K_s$ luminosity-weighted average of the positions of the Milky Way and Andromeda) is located at 
$(+0.12, -0.30, +0.20)$ Mpc.   
The centre of the Council of Giants is at 
$(+0.36, +0.72, 0.00)$ Mpc.  

Missing from the panels of Figure~\ref{sheet_views} are the giants Maffei~1 and 2.   Only their direction is marked (by solid black arrows).
\citet{mcc14a} estimated the distances to be 
$3.3 \pm 0.4 \, \rm Mpc$ 
and $3.4 \pm 0.6 \, \rm Mpc$,  
respectively, founded upon implementations of the Fundamental Plane and the Tully-Fisher relation with independent evaluations of the Galactic extinction from extragalactic probes \citep{fin07a}.  Indeed, purported measurements of the tip of the red giant branch by \citet{wu14a} from HST data for fields near Maffei~1 and Maffei~2 led to consistent distances of 
$3.4 \pm 0.3  \, \rm Mpc$  
and 
$3.5 \pm 0.3 \, \rm Mpc$,  
respectively.
However, re-analysis of the data by \citet{tik18a} increased the distances to 
6.6 and $6.8 \, \rm Mpc$,  
respectively.
Yet another analysis of the same data by \citet{ana19a} concluded that the distance to Maffei 2 is actually
$5.7 \pm 0.4 \, \rm Mpc$.  
It was inferred that Maffei~1 is equally distant.   Besides being at odds with the previous estimates by \citet{mcc14a} for Maffei~1 and Maffei~2  (and Dwingeloo~1, which is in the same area), the new distance shifts the projected peculiar velocity for the Maffei pair from a modest 
$+9  \, \rm km \, s^{-1}$  
(close to that of the adjacent giant IC~342, which is
$-9  \, \rm km \, s^{-1}$)  
to an extreme
$-161 \, \rm km \, s^{-1}$.   
Notably, the estimate of reddening adopted by \citet{ana19a} was founded upon studies of stars in the Milky Way, so it may underestimate the extinction of extragalactic sources (see \citealt{but83a}).

\subsection{Peculiar Velocities}

Any gravitational competition between the Local Group and Council of Giants should be most evident in peculiar motions parallel to the plane of the Local Sheet.  To arrive at them, heliocentric velocities of sample dwarfs were corrected for the Hubble flow and the solar motion with respect to the Council.

To subtract off the component of motion along the line of sight arising from the Hubble flow, namely $H_0 D_{\textit{dw}-\odot}$ where $D_{\textit{dw}-\odot}$ is the distance of a dwarf from the Sun, the local value of the Hubble constant was adopted to be 
$71.6 \pm 2.9 \, \rm  km \, s^{-1} \, Mpc^{-1}$  
(as in \citealt{mcc14a}).
This value is based upon the infrared P-L relation derived for Cepheid variables by \citet{rie11a, rie12a}, but adjusted to the maser distance for M106 \citep{hum13a}.

Resulting line-of-sight peculiar velocities were converted to three-dimensional peculiar velocities relative to the Local Group (LG) by correcting for the motion of the Sun with respect to the luminosity-weighted centroid in $K_s$.  Apices $A = (\left\lvert V \right\rvert,L_\mathit{S},B_\mathit{S})$ of the Sun's speed $\left\lvert V \right\rvert$ (in $\rm km \, s^{-1}$) 
towards longitude and latitude $(L_\mathit{S},B_\mathit{S})$ in the Rotated Sheet Coordinate System were adopted to be
\begin{eqnarray}
A_\textit{$\odot$-LSR} &=& (18 \pm 2, 35.6, +65.3) \nonumber \\  
A_\textit{LSR-MW} &=& (239 \pm 5, 315.5, +43.6) \nonumber \\  
A_\textit{MW-LG} &=& (47 \pm 9, 291.1, +3.4) \nonumber  
\end{eqnarray}
where  $A_\textit{$\odot$-LSR}$ refers to the Sun's motion with respect to the Local Standard of Rest (LSR),  $A_\textit{LSR-MW}$ to the motion of the LSR with respect to the Milky Way (MW), and $A_\textit{MW-LG}$ to the motion of the Milky Way 
with respect to the $K_s$ luminosity-weighted centroid of the Local Group.\citep{sch10a, mcm11a, mar12a, mcc14a}.  

Next a correction was made for the peculiar motion of the Local Group with respect to the Council of Giants \citep{mcc14a}, the apex for which is
\begin{equation}
A_\textit{LG-Council} = (12 \pm 13, 303.7, +1.3)  
\end{equation}
if Maffei 1 and 2 are excluded from the Council.  The apex speed and direction are within 
$1 \, \rm km s^{-1}$  
and $0\fdg3$,  
respectively, of the values determined by \citet{mcc14a} with Maffei 1 and 2 included.
The resulting apex of the motion of the Sun with respect to the Council is
\begin{equation}
\label{eq_solar_motion}
A_\textit{$\odot$-Council} = (294 \pm 9, 311.9, +38.8)  
\end{equation} 

Finally, projections of peculiar velocities along the line of sight from the Sun were determined.  The component parallel to the plane of the Local Sheet, which is the subject of this paper, is given for each galaxy in Table~\ref{mccall_tab1}.  Henceforth, it will be referred to as the {\it projected peculiar velocity}.

\section{Analysis} \label{analysis}

\begin{figure} 
\includegraphics[keepaspectratio=true,width=8.6cm]{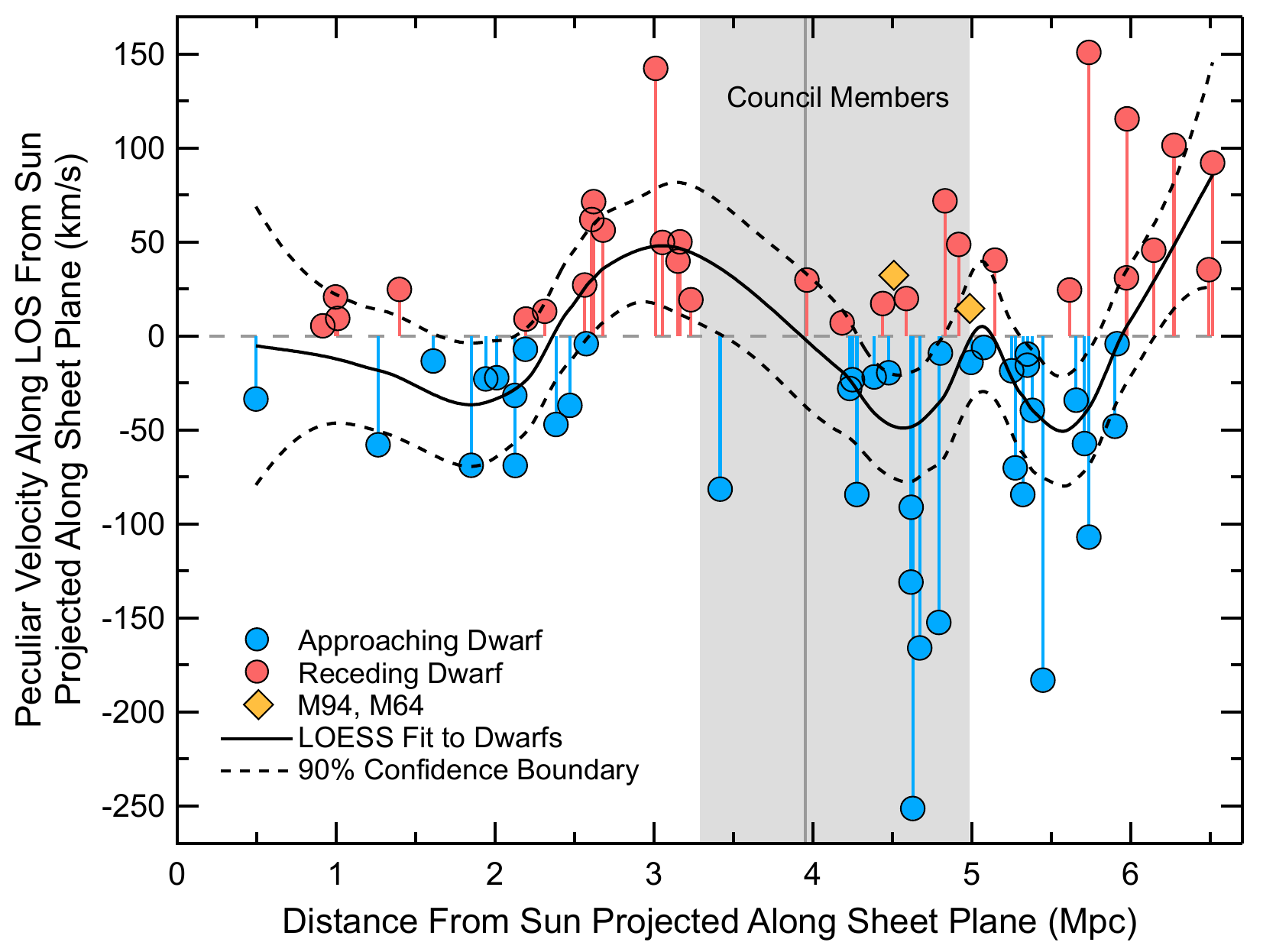}
\caption{\label{vpec_vs_dinplane}
{Projected peculiar velocities of isolated dwarf galaxies versus projected distance from the Sun}.  Each plotted quantity is a measurement along the line of sight (LOS) from the Sun projected parallel to the {plane of the Local Sheet}.  Approaching and receding dwarfs are marked in blue and red, respectively.  Council members span a range of distance that is shaded grey, and their mean distance {(excluding Maffei 1 and 2)} is denoted by the vertical grey line.  {The Council giants M94 and M64 are depicted with yellow diamonds.}  The solid black curve shows the result of LOESS smoothing the dwarf data, and the dashed black curves delineate the {90\%} confidence limits.}
\end{figure}

{Figure~\ref{sheet_views} marks dwarfs with approaching and receding projected peculiar velocities with blue and red symbols, respectively.  Figure~\ref{vpec_vs_dinplane} displays the projected peculiar velocities as a function of projected distance, using the same colours to identify the direction of motion.  In this figure, the domain of Council members is depicted by the shaded grey area and the mean distance, excluding the Maffei galaxies, is marked by a vertical line.}  

Figure~\ref{vpec_vs_dinplane} clearly shows that the motions of dwarfs are not random.  Within
$2.5 \, \rm Mpc$  
of the Milky Way, most dwarfs are approaching.  
Beyond, there are flows towards the realm of Council giants from both sides.  Thus, the Figure provides strong evidence that the Council of Giants is gravitationally disturbing the neighbourhood of the Local Group.

\begin{figure}  
\includegraphics[keepaspectratio=true,width=8.6cm]{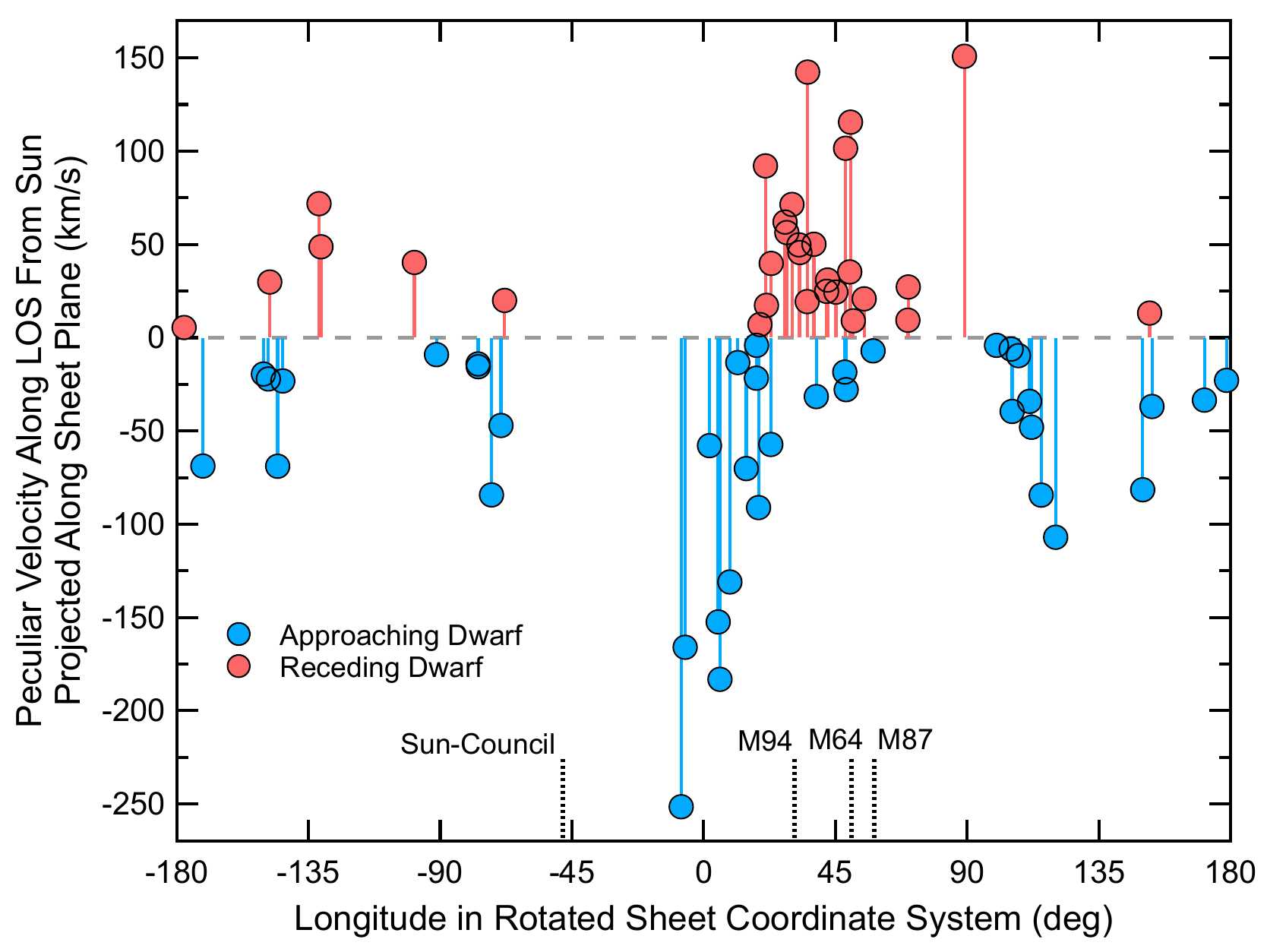}
\caption{\label{vpec_vs_longitude}
Projected peculiar velocities of isolated dwarf galaxies versus longitude in the Rotated Sheet Coordinate System.  The longitude of of the apex of the Sun's motion with respect to the Council of Giants is marked on the abscissa with a short dotted vertical line (labeled Sun-Council).  Displayed in the same way are the directions of the {Council giants M94 and M64 as well as the direction of M87, the centre of the Virgo Cluster.}}
\end{figure}

{
Most galaxies between 
$2.5$  
and 
$4.2 \, \rm Mpc$   
have positive projected peculiar velocities.}
However, they are not randomly placed around the sky.  How peculiar motions are distributed with longitude in the Rotated Sheet Coordinate System is displayed in Figure~\ref{vpec_vs_longitude}, which shows that most of the receding galaxies are in the general direction of M94 (see also Figure~\ref{sheet_views}). 
\noindent
Unfortunately, the range of distance they span is not well sampled in other directions, so it is unclear if this is a local phenomenon or an effect that is more widespread.  The pattern and amplitude of peculiar velocity variations with longitude can't be explained by the uncertainty in the solar motion.  Notably, the longitude of M94 is 
$79^\circ$  
away from that of the apex, so velocity corrections for dwarfs in that general direction are small.

To better establish the locations of inflections in the trend of projected peculiar velocities with projected distance and to evaluate their dependence on the noise, the data were smoothed using robust LOESS curve fitting.  For each point, a local regression was determined by fitting a polynomial of degree 
2  
by least squares to the neighbours in a window spanning 
{
35\%} 
of the data.  More weight was given to neighbours near the considered point and less weight to neighbours farther away.   The selected spanning window delivered a rendition of the data that was neither overly granular nor washed out.  The iterative "robust" version of LOESS smoothing was employed to reduce the sensitivity to outliers.  The resulting curve and 
{
90\%} 
confidence limits are superimposed on Figure~\ref{vpec_vs_dinplane}.

Moving outward from the Sun, there is a steep transition from negative to positive peculiar velocities on the inner side of the Council at a radius of 
{
$2.36 \pm 0.23 \, \rm Mpc$},  
where the uncertainty is derived from the 
{
90\%} 
confidence boundaries of the LOESS fit.   In comparison, through a process of density matching, \citet{mcc14a} estimated 
the ``edge'' of the Local Group to be at a radius of 
$2.56 \pm 0.17 \, \rm Mpc$  
($2.68 \pm 0.19 \, \rm Mpc$  
if Maffei 1 and 2 are excluded).  Farther out, there is a shallower transition from positive to
{
mostly}
negative velocities at 
{
$3.93^{+0.35}_{-0.49} \, \rm Mpc$}. 
In comparison, the radius of the Council is 
$3.75 \pm 0.20 \, \rm Mpc$.  
The inner transition will be referred to as the {\it ascending node} and the outer transition as the {\it descending node}.   
If the Hubble constant is changed by an amount consistent with its uncertainty 
($ \pm 2.9 \, \rm  km \, s^{-1} Mpc^{-1}$),  
the positions of the nodes shift by amounts well within the quoted errors.   

{
A model of flows within $38 \, \rm Mpc$ was constructed by \citet{sha17a} by using the numerical action method to arrive at gravitationally-induced trajectories for galaxies having reliable distances in the 
CosmicFlows-3 Database \citep{tul16a}, which is attached to the EDD.   Resulting peculiar motions today proved to be governed primarily by a flow away from the Local Void down to the supergalactic plane and a flow along the supergalactic plane towards the Virgo Cluster.  The flow towards Virgo should be manifested in the heliocentric velocities of the sample of galaxies studied here because the plane of the Local Sheet is inclined by only $8^\circ$ to the Local Supercluster.  However, it should be suppressed in a reference frame defined by Council giants.  

To explore flow predictions for the realm of the Council of Giants, the Cosmicflows-3 Distance-Velocity Calculator\footnote{https://edd.ifa.hawaii.edu/NAMcalculator/} included in the EDD, which is the interactive implementation of the model \citep{kou20a},  was utilized.
Input distances were adjusted for the difference in Hubble constants so as to preserve predicted deviations from the Hubble flow. Projected peculiar velocities with respect to the Council were determined from predicted heliocentric velocities using the apex of the solar motion given by equation~\ref{eq_solar_motion}.

For Council giants M64 and M94, whose longitudes are respectively $8^\circ$ and $27^\circ$ away from the direction of the Virgo cluster (see Figures~\ref{sheet_views} and \ref{vpec_vs_longitude}), the observed projected peculiar velocities in the Council frame are 
15 and $32 \, \rm km \, s^{-1}$, 
respectively (see Figure~\ref{vpec_vs_dinplane}).  However, the model substantially over-estimates them, yielding 
88 and $58 \, \rm km \, s^{-1}$, 
respectively.  
Yet, the dwarf IC~3840, which lies in the direction of M64 but 
$1 \, \rm Mpc$
beyond, is predicted to be receding $30 \, \rm km \, s^{-1}$ slower than its actual velocity of 
$116 \, \rm km \, s^{-1}$.
Also, the model significantly under-estimates projected peculiar velocities for receding dwarfs in the direction of but nearer than M94; observed values for NGC 3741, DDO 99, and DDO 147 are 40, 71, and 
$142 \, \rm km \, s^{-1}$,
respectively, whereas predicted values are
0, 18, and $48 \, \rm km \, s^{-1}$. 
But for NGC~3738, which lies
beyond the Council radius at a longitude $16^\circ$
less than that of M94, the predicted peculiar velocity is 
$70 \, \rm km \, s^{-1}$ more negative than the observed velocity of 
$-70 \, \rm km \, s^{-1}$.  In the second, third, and fourth quadrants, however, predicted peculiar motions for approaching dwarfs beyond the Council radius are too slow.  
In summary, the flow model does not deliver very accurate velocities numerically for individual galaxies in the domain of the Council of Giants, although it generally gives the correct signs.

The low projected peculiar velocity observed for M64 is notable because the galaxy lies so close to the line of sight to M87.  This indicates that any residual of the Virgo flow must be small.  Even though M64 is nearly at rest in the Council frame, it is conceivable that the recessional motions of dwarfs a megaparsec or more beyond are attributable in part to the flow towards Virgo.  However, Figure~\ref{sheet_views} does not reveal any widespread imprint of such a flow; there are dwarfs beyond the Council radius in all directions that are seen to be moving inward.  Although M94 is receding, nearer receding dwarfs in its direction are moving fast enough to be approaching it.}

\section{Discussion} \label{discussion}

Flows are a natural consequence of the existence of mass concentrations, so one might argue that flows towards the Local Group and Council members need not be coupled.  
However, if local motions were a consequence of independent flows, then there ought to be a zone between the Local Group and the Council (or its consituents) where there is no preferred flow direction.  On the contrary, with increasing distance projected peculiar velocities smoothly transition from negative to positive and back to negative.

Given how dwarfs with positive peculiar velocities are distributed in longitude, a possible explanation for the ascending node is a localized tug of war between the Local Group and M94.  If correct, the ascending node would be located at the distance $r_a$ from the centroid of the Local Group at which gravitational forces balance, in which case the ratio of the total mass $\mathcal{M}_\textit{M94}$ of M94 relative to the total mass $\mathcal{M}_\textit{LG}$ of the Local Group would be given by
\begin{equation}
\mathcal{M}_\textit{M94} / \mathcal{M}_\textit{LG} = \left[ (r_\textit{M94} / r_a) -1 \right]^2  
\end{equation}
where $r_\textit{M94}$ is the distance of M94 from the centroid of the Local Group.  From \citet{mcc14a}, the heliocentric distances of M94 and the centroid of the Local Group are 
$4.51 \pm 0.11 \, \rm Mpc$  
and 
$0.33 \pm 0.03 \, \rm Mpc$,  
respectively.  Thus, $r_\textit{M94}$ is 
$4.57 \pm 0.11 \, \rm Mpc$.   
In the direction of M94, $r_a$ is 
{
$2.41 \pm 0.23 \, \rm Mpc$}. 
Consequently, 
{
$\mathcal{M}_\textit{M94} / \mathcal{M}_\textit{LG} = 0.80^{+0.42}_{-0.27}$.
Where the attraction of the Local Group balances that of M94 is marked with a solid black circle in Figure~\ref{sheet_views}.} 

{
Stellar masses yield a value for  $\mathcal{M}_\textit{M94} / \mathcal{M}_\textit{LG}$ of only 
$0.12 \pm 0.02$  
if the masses of dark matter haloes scale proportionately (see \citealt{mcc14a}).  
However, orbital motions of satellites reveal that there may be seven times times more matter per unit stellar mass associated with M94 compared to M31 and the Milky Way, and that the total mass relative to the Local Group is actually $0.85 \pm 0.32$ \citep{kar21a}.  The estimate should account for most of the matter in the Canes Venatici I Cloud, as the next largest galaxy, NGC 4449, has a stellar mass only 
7\%  
of that of M94 \citep{mcc14a}.
Although uncertainties are large, the remarkable agreement supports the conjecture that the ascending node represents a transition between two gravitational realms.}

Another scenario for explaining the nodes is a matter distribution that is widely distributed in longitude, i.e., that is more Council-like.  It is motivated by the wide range of longitude over which isolated dwarfs are observed to be approaching.  In this case, it is informative to model the distribution as a thin torus with a uniform linear density and a radius $R$ equal to that of the Council of Giants, i.e., 
$3.75 \pm 0.20 \, \rm Mpc$.  
Projected onto the Local Sheet, the centroid of the Local Group is offset from the centre of the Council by 
$1.0 \, \rm Mpc$.
Consequently, the locus of points where the gravity of the Local Group balances the net radial pull of the torus is noncircular.   
{The location of the ascending node should be close to the mean distance of these locations from the Sun.}
Adopting 
$r_a / R = 0.63 \pm 0.07$,  
then the total mass of the torus $\mathcal{M}_\textit{torus}$ relative to the mass of the Local Group must be
{$\mathcal{M}_\textit{torus} / \mathcal{M}_\textit{LG} = 4.3^{+3.0}_{-1.8}$. 
The corresponding locus of balance points is indicated by a black dashed ellipse (ellipticity 0.97) in Figure~\ref{sheet_views}.}

{
An independent estimate of $\mathcal{M}_\textit{torus} / \mathcal{M}_\textit{LG}$ can be gained from masses derived from the orbits of satellites around individual members of the Council.  Such data are available for six of those galaxies \citep{kar21a}.
Based upon the fully-corrected $K_s$ luminosities of four of the five disk galaxies in the set \citep{mcc14a}, the corresponding average mass-to-light ratio is $19 \pm 12 \, \mathcal{M}_\odot / \mathcal{L}_\odot$, where the uncertainty is the standard deviation (M94 was excluded because it is an outlier).  Despite the different treatment of $K_s$ data, the result is close to the average for disk galaxies found by \citet{kar21a}.  Applying it to the luminosities of the four remaining disk galaxies lacking satellite measurements, the total mass of the Council relative to the Local Group comes out to be $6.3 \pm 1.4$.   The ratio rises to $7.6 \pm 1.7$  if Maffei~1 and 2 are included at the distances determined by \citet{mcc14a} (adopting the mass-to-light ratio of NGC~5128 for the former).  Within errors, the result from the ascending node can be considered to be in agreement with that based on satellite orbits, lending some credence to the idea that the ascending node marks where the gravity of the Local Group balances pulls from more than one direction.}

A limitation of this study is lack of knowledge about tangential velocities, so it has not been possible to explore with any confidence how the distribution of projected peculiar velocities with projected distance changes with the choice of origin.  The Sun's velocity with respect to the Council is known in three dimensions, but the heliocentric velocity of each dwarf is one-dimensional.  A change of origin that leads to a shift in longitude delivers a line-of-sight velocity that is diminished from the heliocentric value yet which is fully corrected for the solar motion in that direction.  Consequently, any trends that may exist can be washed out by the loss of information.

\section{Conclusions} \label{conclusions}

Motions of isolated dwarf galaxies in the Local Sheet display evidence for flows towards both the Local Group and the Council of Giants.  Heliocentric peculiar velocities in the frame of reference of the Council and projected along the Local Sheet transition from negative to positive and positive to negative at heliocentric distances of 
$2.4 \pm 0.2 \, \rm Mpc$ 
(the ascending node)
and 
{
$3.9^{+0.4}_{-0.5} \, \rm Mpc$} 
(the descending node), 
respectively, projected along the plane of the Sheet.   Within errors, the former is the same as the gravitational range of the Local Group judged from density matching and the latter is close to the centre of the domain of Council members.  Receding dwarfs on the near side of the Council are primarily located in the general direction of M94, although in part due to the limitations of sampling.  If the ascending node is attributed to a localized balance of gravitational forces between M94 and the Local Group, the implied total mass of M94 relative to the Local Group is 
{
$0.80^{+0.42}_{-0.27}$,  
essentially the same as the ratio 
$0.85\pm 0.32$ estimated from the orbits of satellite galaxies.}
If instead the ascending node is a manifestation of matter widely spread in longitude, as the longitudinal distribution of approaching dwarfs suggests, then the required ratio of the gravitating mass relative to the Local Group is 
$4.3^{+3.0}_{-1.8}$. 
{
In comparison, masses for Council and Local Group giants estimated from the motions of satellite galaxies yield the ratio
$6.3 \pm 1.4$ if the Maffei galaxies are excluded and 
$7.6 \pm 1.7$ if they aren't.  Although the agreement is poorer than the outcome of the M94 analysis, the results can still be considered to be consistent within errors.}
{
The compatibility of predictions for mass ratios with observations suggests that the proposed interpretation of nodes is correct and that the gravitational sphere of influence of the Local Group is truncated by the Council of Giants.}

\section*{Acknowledgements}
MLM thanks G.~J. Conidis for his suggestion to use dwarfs to seek Council-driven flows, 
{
and to I. Karachentsev for bringing attention to modern resources invaluable to sample selection and the interpretation of results.}
MLM is particularly grateful to B. Nathoo, F. Shariff, and J. Vecchiarelli for their dedication to preserving his health, and to M. Martin for moral support.  York University is thanked for its continuing financial support.  This publication has made use of the software Matlab, Mathematica, and IGOR PRO.

\section*{Data Availability Statement}
The data underlying this article were accessed from the cited references as well as the {
Catalog \& Atlas of the LV Galaxies (https://www.sao.ru/lv/lvgdb/),
HyperLeda (http://leda.univ-lyon1.fr/),
The Extragalactic Distance Database (http://edd.ifa.hawaii.edu/dfirst.php), 
the Cosmicflows-3 Distance-Velocity Calculator (https://edd.ifa.hawaii.edu/NAMcalculator/),
NED (http://ned.ipac.caltech.edu),
and SIMBAD (https://simbad.u-strasbg.fr/simbad/).}
The derived data generated in this research will be shared on reasonable request to the corresponding author. 

\bibliographystyle{mnras}
\bibliography{mccall_references}


\clearpage
\onecolumn
\setcounter{table}{0}

\begin{center}
\begin{ThreePartTable}

\begin{TableNotes}

\item[]
(1) Name of galaxy in the Local Volume Galaxy Database (LVG), in order of $d_\parallel$;
(2) Supergalactic longitude ({HyperLeda});
(3) Supergalactic latitude ({HyperLeda});
(4) Distance Modulus, in mag (see text);
(5) Heliocentric radial velocity, in $\rm km \, s^{-1}$ ({HyperLeda});
(6),(7),(8) Cartesian coordinates in the Rotated Sheet Coordinate System, in Mpc (as defined by \citealt{mcc14a}). The origin is the projection of the Sun onto the plane of the Local Sheet;
(9) Longitude in the Rotated Sheet Coordinate System, in deg;
(10) Distance from the Sun projected onto the plane of the Local Sheet, in Mpc { (the {\it projected distance})};
(11) Peculiar line-of-sight velocity in the frame of reference of the Council of Giants but projected onto the plane of the Local Sheet, in $\rm km \, s^{-1}$ { (the {\it projected peculiar velocity})}.


\end{TableNotes}

\setlength{\jot}{-0.0pt}

\begin{longtable}{lrrcrrrrrrr}

\caption{
Dwarf Galaxy Sample
\label{mccall_tab1}
}
\\

\hline
\noalign{\smallskip}


Galaxy &
$L$~~~ &
$B$~~ &
$\mathit{DM}$~ &
$V_\odot$~~~~~~ &
$X_\mathit{S}$~~ &
$Y_\mathit{S}$~~ &
$Z_\mathit{S}$~~ &
$L_\mathit{S}$~~ &
$d_\parallel$~~ &
$V_{\parallel, \mathit{pec}}$
\\



(1)~~&
(2)~~ &
(3)~ &
(4)~~ &
(5)~~~~~~ &
(6)~~ &
(7)~~ &
(8)~~ &
(9)~~ &
(10)~ &
(11)~
\\

\noalign{\smallskip}
\hline
\noalign{\smallskip}

\endfirsthead

\caption{cont'd.} \\

\hline
\noalign{\smallskip}


Galaxy &
$L$~~~ &
$B$~~ &
$\mathit{DM}$~ &
$V_\odot$~~~~~~ &
$X_\mathit{S}$~~ &
$Y_\mathit{S}$~~ &
$Z_\mathit{S}$~~ &
$L_\mathit{S}$~~ &
$d_\parallel$~~ &
$V_{\parallel, \mathit{pec}}$
\\



(1)~~ &
(2)~~ &
(3)~ &
(4)~~ &
(5)~~~~~~ &
(6)~~ &
(7)~~ &
(8)~~ &
(9) ~~&
(10)~ &
(11)~
\\

\noalign{\smallskip}
\hline
\endhead

\noalign{\smallskip}
\hline

\endfoot
\hline
\noalign{\smallskip}
\insertTableNotes

\endlastfoot


Sag dIr	&	$	221.3	$	&	$	55.5	$	&	$	25.17	\pm	0.08	$	&	$	-78.7	\pm	1.9	$	&	$	-0.49	$	&	$	0.08	$	&	$	1.09	$	&	$	171.0	$	&	0.49	&	$	-33.5	$	\\
Tucana	&	$	227.6	$	&	$	-0.9	$	&	$	24.82	\pm	0.04	$	&	$	194.0	\pm	4.3	$	&	$	-0.91	$	&	$	-0.04	$	&	$	0.24	$	&	$	-177.5	$	&	0.92	&	$	5.5	$	\\
SexB	&	$	95.5	$	&	$	-39.6	$	&	$	25.78	\pm	0.03	$	&	$	301.0	\pm	0.6	$	&	$	0.57	$	&	$	0.81	$	&	$	-0.90	$	&	$	54.8	$	&	0.99	&	$	20.9	$	\\
SexA	&	$	109.0	$	&	$	-40.7	$	&	$	25.80	\pm	0.08	$	&	$	323.9	\pm	0.7	$	&	$	0.35	$	&	$	0.95	$	&	$	-0.91	$	&	$	69.7	$	&	1.01	&	$	9.4	$	\\
UGC04879	&	$	47.6	$	&	$	-15.0	$	&	$	25.68	\pm	0.03	$	&	$	-29.1	\pm	0.9	$	&	$	1.26	$	&	$	0.04	$	&	$	-0.40	$	&	$	1.9	$	&	1.26	&	$	-57.8	$	\\
LeoP	&	$	84.9	$	&	$	-28.9	$	&	$	26.19	\pm	0.33	$	&	$	263.7	\pm	3.0	$	&	$	1.04	$	&	$	0.93	$	&	$	-0.89	$	&	$	41.9	$	&	1.40	&	$	24.9	$	\\
KKR25	&	$	56.1	$	&	$	40.4	$	&	$	26.40	\pm	0.07	$	&	$	-75.3	\pm	9.9	$	&	$	1.58	$	&	$	0.32	$	&	$	1.15	$	&	$	11.6	$	&	1.61	&	$	-13.3	$	\\
IC5152	&	$	234.2	$	&	$	11.5	$	&	$	26.46	\pm	0.04	$	&	$	123.8	\pm	1.9	$	&	$	-1.83	$	&	$	-0.29	$	&	$	0.78	$	&	$	-171.1	$	&	1.85	&	$	-68.8	$	\\
\lbrack KK2000 \rbrack\ 03	&	$	222.8	$	&	$	-21.3	$	&	$	26.50	\pm	0.09	$	&	$	316.0	\pm	7.0	$	&	$	-1.94	$	&	$	0.05	$	&	$	-0.35	$	&	$	178.6	$	&	1.94	&	$	-22.8	$	\\
KK258	&	$	255.5	$	&	$	18.6	$	&	$	26.75	\pm	0.03	$	&	$	92.0	\pm	5.0	$	&	$	-1.72	$	&	$	-1.04	$	&	$	1.12	$	&	$	-148.7	$	&	2.01	&	$	-22.1	$	\\
KK230	&	$	84.6	$	&	$	23.5	$	&	$	26.72	\pm	0.22	$	&	$	61.0	\pm	2.1	$	&	$	1.66	$	&	$	1.32	$	&	$	0.75	$	&	$	38.5	$	&	2.12	&	$	-31.6	$	\\
UGCA438	&	$	258.9	$	&	$	9.3	$	&	$	26.73	\pm	0.08	$	&	$	62.0	\pm	2.6	$	&	$	-1.75	$	&	$	-1.20	$	&	$	0.77	$	&	$	-145.6	$	&	2.13	&	$	-68.9	$	\\
GR8	&	$	103.0	$	&	$	4.7	$	&	$	26.70	\pm	0.11	$	&	$	217.0	\pm	2.1	$	&	$	1.17	$	&	$	1.85	$	&	$	0.08	$	&	$	57.8	$	&	2.19	&	$	-7.1	$	\\
DDO187	&	$	97.8	$	&	$	24.4	$	&	$	26.81	\pm	0.04	$	&	$	152.9	\pm	1.3	$	&	$	1.38	$	&	$	1.71	$	&	$	0.84	$	&	$	51.1	$	&	2.19	&	$	9.1	$	\\
IC3104	&	$	195.8	$	&	$	-17.1	$	&	$	26.86	\pm	0.02	$	&	$	429.4	\pm	3.2	$	&	$	-2.05	$	&	$	1.08	$	&	$	-0.34	$	&	$	152.3	$	&	2.31	&	$	13.2	$	\\
KKH98	&	$	332.4	$	&	$	23.2	$	&	$	27.06	\pm	0.09	$	&	$	-134.3	\pm	2.0	$	&	$	0.84	$	&	$	-2.23	$	&	$	1.13	$	&	$	-69.3	$	&	2.38	&	$	-47.0	$	\\
IC4662	&	$	199.2	$	&	$	8.6	$	&	$	27.03	\pm	0.02	$	&	$	305.5	\pm	3.2	$	&	$	-2.20	$	&	$	1.12	$	&	$	0.76	$	&	$	153.1	$	&	2.47	&	$	-36.9	$	\\
KKH86	&	$	116.3	$	&	$	15.5	$	&	$	27.08	\pm	0.26	$	&	$	285.7	\pm	1.9	$	&	$	0.88	$	&	$	2.41	$	&	$	0.62	$	&	$	69.8	$	&	2.56	&	$	27.2	$	\\
UGC08508	&	$	63.1	$	&	$	17.9	$	&	$	27.08	\pm	0.04	$	&	$	59.9	\pm	1.3	$	&	$	2.45	$	&	$	0.80	$	&	$	0.58	$	&	$	18.0	$	&	2.57	&	$	-4.1	$	\\
DDO125	&	$	72.8	$	&	$	5.9	$	&	$	27.08	\pm	0.05	$	&	$	199.3	\pm	1.3	$	&	$	2.31	$	&	$	1.22	$	&	$	0.04	$	&	$	27.8	$	&	2.61	&	$	62.1	$	\\
DDO099	&	$	74.9	$	&	$	-2.1	$	&	$	27.12	\pm	0.08	$	&	$	248.5	\pm	1.4	$	&	$	2.26	$	&	$	1.31	$	&	$	-0.33	$	&	$	30.1	$	&	2.62	&	$	71.5	$	\\
DDO190	&	$	74.1	$	&	$	26.9	$	&	$	27.26	\pm	0.04	$	&	$	151.9	\pm	1.5	$	&	$	2.36	$	&	$	1.27	$	&	$	1.06	$	&	$	28.4	$	&	2.68	&	$	56.4	$	\\
DDO147	&	$	80.6	$	&	$	7.7	$	&	$	27.39	\pm	0.08	$	&	$	331.9	\pm	1.9	$	&	$	2.45	$	&	$	1.75	$	&	$	0.14	$	&	$	35.5	$	&	3.01	&	$	142.5	$	\\
DDO181	&	$	78.1	$	&	$	18.6	$	&	$	27.46	\pm	0.04	$	&	$	201.8	\pm	1.5	$	&	$	2.57	$	&	$	1.64	$	&	$	0.72	$	&	$	32.5	$	&	3.05	&	$	49.8	$	\\
NGC3741	&	$	68.0	$	&	$	-2.1	$	&	$	27.52	\pm	0.04	$	&	$	228.7	\pm	0.8	$	&	$	2.90	$	&	$	1.23	$	&	$	-0.43	$	&	$	23.0	$	&	3.15	&	$	39.9	$	\\
UGC08833	&	$	83.5	$	&	$	21.1	$	&	$	27.56	\pm	0.06	$	&	$	224.7	\pm	2.0	$	&	$	2.50	$	&	$	1.93	$	&	$	0.90	$	&	$	37.6	$	&	3.16	&	$	50.2	$	\\
DDO183	&	$	81.1	$	&	$	20.5	$	&	$	27.60	\pm	0.05	$	&	$	190.7	\pm	1.9	$	&	$	2.63	$	&	$	1.87	$	&	$	0.87	$	&	$	35.4	$	&	3.23	&	$	19.4	$	\\
HIPASS J1247-77	&	$	193.5	$	&	$	-15.7	$	&	$	27.70	\pm	0.17	$	&	$	413.2	\pm	3.6	$	&	$	-2.95	$	&	$	1.71	$	&	$	-0.50	$	&	$	149.9	$	&	3.41	&	$	-81.5	$	\\
NGC0625	&	$	257.3	$	&	$	-17.7	$	&	$	28.02	\pm	0.04	$	&	$	390.3	\pm	6.5	$	&	$	-3.37	$	&	$	-2.08	$	&	$	-0.57	$	&	$	-148.3	$	&	3.96	&	$	29.8	$	\\
UGC06541	&	$	64.2	$	&	$	-0.8	$	&	$	28.13	\pm	0.12	$	&	$	251.0	\pm	1.7	$	&	$	3.95	$	&	$	1.38	$	&	$	-0.51	$	&	$	19.2	$	&	4.18	&	$	7.0	$	\\
UGC08638	&	$	94.7	$	&	$	16.3	$	&	$	28.16	\pm	0.03	$	&	$	275.0	\pm	0.7	$	&	$	2.79	$	&	$	3.18	$	&	$	0.85	$	&	$	48.7	$	&	4.23	&	$	-27.8	$	\\
UGCA442	&	$	260.8	$	&	$	6.1	$	&	$	28.20	\pm	0.08	$	&	$	267.6	\pm	5.9	$	&	$	-3.43	$	&	$	-2.51	$	&	$	1.16	$	&	$	-143.8	$	&	4.25	&	$	-23.2	$	\\
UGC01703	&	$	333.5	$	&	$	-7.1	$	&	$	28.17	\pm	0.03	$	&	$	40.0	\pm	20.0	$	&	$	1.28	$	&	$	-4.08	$	&	$	-0.41	$	&	$	-72.5	$	&	4.28	&	$	-84.3	$	\\
NGC4068	&	$	62.9	$	&	$	4.9	$	&	$	28.21	\pm	0.02	$	&	$	209.6	\pm	1.7	$	&	$	4.17	$	&	$	1.35	$	&	$	-0.10	$	&	$	17.9	$	&	4.38	&	$	-21.7	$	\\
NGC5238	&	$	66.6	$	&	$	18.4	$	&	$	28.27	\pm	0.02	$	&	$	228.7	\pm	0.9	$	&	$	4.13	$	&	$	1.62	$	&	$	0.95	$	&	$	21.4	$	&	4.44	&	$	17.3	$	\\
ESO245-005	&	$	255.1	$	&	$	-19.7	$	&	$	28.30	\pm	0.06	$	&	$	393.7	\pm	2.3	$	&	$	-3.89	$	&	$	-2.21	$	&	$	-0.82	$	&	$	-150.4	$	&	4.48	&	$	-19.5	$	\\
KKH18	&	$	339.3	$	&	$	-15.9	$	&	$	28.40	\pm	0.13	$	&	$	211.3	\pm	4.6	$	&	$	1.71	$	&	$	-4.25	$	&	$	-1.26	$	&	$	-68.1	$	&	4.59	&	$	20.0	$	\\
UGC06757	&	$	53.9	$	&	$	5.9	$	&	$	28.32	\pm	0.14	$	&	$	81.8	\pm	2.7	$	&	$	4.56	$	&	$	0.72	$	&	$	-0.03	$	&	$	8.9	$	&	4.62	&	$	-130.9	$	\\
MCG +09-20-131	&	$	63.7	$	&	$	6.6	$	&	$	28.32	\pm	0.09	$	&	$	154.0	\pm	0.8	$	&	$	4.37	$	&	$	1.48	$	&	$	0.02	$	&	$	18.7	$	&	4.62	&	$	-91.1	$	\\
UGC06456	&	$	36.9	$	&	$	11.4	$	&	$	28.33	\pm	0.04	$	&	$	-107.0	\pm	8.2	$	&	$	4.59	$	&	$	-0.62	$	&	$	0.47	$	&	$	-7.7	$	&	4.63	&	$	-251.4	$	\\
UGC08245	&	$	37.9	$	&	$	16.1	$	&	$	28.37	\pm	0.03	$	&	$	-26.3	\pm	4.7	$	&	$	4.64	$	&	$	-0.52	$	&	$	0.85	$	&	$	-6.4	$	&	4.67	&	$	-165.9	$	\\
DDO165	&	$	49.6	$	&	$	15.6	$	&	$	28.42	\pm	0.02	$	&	$	31.2	\pm	2.8	$	&	$	4.77	$	&	$	0.41	$	&	$	0.79	$	&	$	5.0	$	&	4.79	&	$	-152.4	$	\\
UGC00685	&	$	313.3	$	&	$	1.6	$	&	$	28.41	\pm	0.02	$	&	$	155.9	\pm	2.5	$	&	$	-0.11	$	&	$	-4.80	$	&	$	0.47	$	&	$	-91.3	$	&	4.80	&	$	-9.1	$	\\
NGC0059	&	$	273.1	$	&	$	3.2	$	&	$	28.45	\pm	0.03	$	&	$	366.0	\pm	3.1	$	&	$	-3.19	$	&	$	-3.62	$	&	$	0.97	$	&	$	-131.4	$	&	4.83	&	$	71.9	$	\\
DDO226	&	$	274.2	$	&	$	-3.2	$	&	$	28.46	\pm	0.13	$	&	$	361.1	\pm	1.8	$	&	$	-3.21	$	&	$	-3.72	$	&	$	0.43	$	&	$	-130.8	$	&	4.92	&	$	48.7	$	\\
KK17	&	$	328.7	$	&	$	-6.1	$	&	$	28.50	\pm	0.05	$	&	$	163.9	\pm	2.0	$	&	$	1.11	$	&	$	-4.87	$	&	$	-0.36	$	&	$	-77.1	$	&	4.99	&	$	-14.0	$	\\
ESO379-007	&	$	146.9	$	&	$	-21.0	$	&	$	28.68	\pm	0.07	$	&	$	641.5	\pm	2.3	$	&	$	-1.31	$	&	$	4.90	$	&	$	-1.86	$	&	$	105.0	$	&	5.07	&	$	-6.1	$	\\
PiscesA	&	$	304.3	$	&	$	12.5	$	&	$	28.64	\pm	0.10	$	&	$	235.1	\pm	2.0	$	&	$	-0.79	$	&	$	-5.08	$	&	$	1.61	$	&	$	-98.8	$	&	5.14	&	$	40.4	$	\\
AGC238890	&	$	94.0	$	&	$	14.9	$	&	$	28.62	\pm	0.09	$	&	$	360.0	\pm	3.8	$	&	$	3.50	$	&	$	3.91	$	&	$	0.88	$	&	$	48.2	$	&	5.25	&	$	-18.4	$	\\
NGC3738	&	$	59.6	$	&	$	1.8	$	&	$	28.62	\pm	0.02	$	&	$	224.5	\pm	7.6	$	&	$	5.10	$	&	$	1.32	$	&	$	-0.44	$	&	$	14.6	$	&	5.27	&	$	-70.1	$	\\
NGC5408	&	$	160.6	$	&	$	1.9	$	&	$	28.63	\pm	0.07	$	&	$	506.3	\pm	3.4	$	&	$	-2.27	$	&	$	4.81	$	&	$	0.42	$	&	$	115.2	$	&	5.32	&	$	-84.3	$	\\
ESO381-018	&	$	150.9	$	&	$	-11.2	$	&	$	28.68	\pm	0.05	$	&	$	621.2	\pm	7.2	$	&	$	-1.61	$	&	$	5.10	$	&	$	-0.93	$	&	$	107.5	$	&	5.35	&	$	-9.5	$	\\
NGC0784	&	$	328.8	$	&	$	-6.3	$	&	$	28.65	\pm	0.01	$	&	$	188.8	\pm	4.2	$	&	$	1.20	$	&	$	-5.21	$	&	$	-0.42	$	&	$	-77.0	$	&	5.35	&	$	-15.5	$	\\
ESO381-020	&	$	148.9	$	&	$	-10.5	$	&	$	28.69	\pm	0.05	$	&	$	587.9	\pm	2.6	$	&	$	-1.43	$	&	$	5.19	$	&	$	-0.90	$	&	$	105.4	$	&	5.38	&	$	-39.5	$	\\
UGC07242	&	$	50.4	$	&	$	10.3	$	&	$	28.68	\pm	0.03	$	&	$	63.6	\pm	2.4	$	&	$	5.42	$	&	$	0.53	$	&	$	0.36	$	&	$	5.5	$	&	5.45	&	$	-183.1	$	\\
AGC749241	&	$	90.1	$	&	$	3.7	$	&	$	28.75	\pm	0.06	$	&	$	451.0	\pm	3.3	$	&	$	3.96	$	&	$	3.98	$	&	$	-0.20	$	&	$	45.1	$	&	5.61	&	$	24.5	$	\\
HIPASS J1348-37	&	$	156.4	$	&	$	0.5	$	&	$	28.76	\pm	0.07	$	&	$	581.6	\pm	5.5	$	&	$	-2.05	$	&	$	5.27	$	&	$	0.24	$	&	$	111.3	$	&	5.65	&	$	-34.1	$	\\
UGCA281	&	$	67.9	$	&	$	7.1	$	&	$	28.78	\pm	0.05	$	&	$	279.9	\pm	2.8	$	&	$	5.26	$	&	$	2.22	$	&	$	0.05	$	&	$	22.9	$	&	5.71	&	$	-57.2	$	\\
HIPASS J1351-47	&	$	165.0	$	&	$	-2.2	$	&	$	28.79	\pm	0.12	$	&	$	529.2	\pm	5.3	$	&	$	-2.88	$	&	$	4.96	$	&	$	0.09	$	&	$	120.2	$	&	5.73	&	$	-107.0	$	\\
UGCA319	&	$	133.5	$	&	$	-2.9	$	&	$	28.80	\pm	0.07	$	&	$	748.9	\pm	2.0	$	&	$	0.09	$	&	$	5.73	$	&	$	-0.41	$	&	$	89.1	$	&	5.74	&	$	150.9	$	\\
KK182	&	$	155.9	$	&	$	-8.2	$	&	$	28.87	\pm	0.13	$	&	$	615.4	\pm	2.3	$	&	$	-2.21	$	&	$	5.47	$	&	$	-0.65	$	&	$	112.0	$	&	5.90	&	$	-48.0	$	\\
ESO443-009	&	$	143.9	$	&	$	-7.4	$	&	$	28.88	\pm	0.11	$	&	$	643.4	\pm	3.7	$	&	$	-1.02	$	&	$	5.82	$	&	$	-0.75	$	&	$	100.0	$	&	5.91	&	$	-4.1	$	\\
KK144	&	$	87.1	$	&	$	1.2	$	&	$	28.89	\pm	0.08	$	&	$	482.1	\pm	2.0	$	&	$	4.42	$	&	$	4.02	$	&	$	-0.50	$	&	$	42.3	$	&	5.97	&	$	31.0	$	\\
IC3840	&	$	95.3	$	&	$	5.0	$	&	$	28.88	\pm	0.08	$	&	$	581.8	\pm	2.0	$	&	$	3.83	$	&	$	4.59	$	&	$	-0.04	$	&	$	50.1	$	&	5.97	&	$	115.6	$	\\
Arp211	&	$	77.9	$	&	$	6.4	$	&	$	28.94	\pm	0.08	$	&	$	453.8	\pm	4.6	$	&	$	5.16	$	&	$	3.33	$	&	$	0.00	$	&	$	32.8	$	&	6.14	&	$	45.7	$	\\
KKH80	&	$	93.2	$	&	$	-0.1	$	&	$	29.00	\pm	0.09	$	&	$	602.9	\pm	1.2	$	&	$	4.16	$	&	$	4.69	$	&	$	-0.63	$	&	$	48.5	$	&	6.27	&	$	101.5	$	\\
LV J1249+2155	&	$	95.0	$	&	$	4.6	$	&	$	29.06	\pm	0.10	$	&	$	538.9	\pm	3.1	$	&	$	4.18	$	&	$	4.96	$	&	$	-0.11	$	&	$	49.9	$	&	6.49	&	$	35.3	$	\\
NGC4707	&	$	66.2	$	&	$	11.2	$	&	$	29.07	\pm	0.10	$	&	$	468.0	\pm	0.7	$	&	$	6.08	$	&	$	2.34	$	&	$	0.50	$	&	$	21.1	$	&	6.51	&	$	92.1	$	\\

\noalign{\smallskip}

\end{longtable}

\end{ThreePartTable}

\end{center}


\bsp	
\label{lastpage}
\end{document}